\documentstyle[epsfig,12pt]{article}
\catcode`\@=11 
\let\rel@x=\relax
\newcount\timecount
\newcount\hours \newcount\minutes  \newcount\temp \newcount\pmhours

\hours = \time
\divide\hours by 60
\temp = \hours
\multiply\temp by 60
\minutes = \time
\advance\minutes by -\temp
\def\hour{\the\hours}
\def\minute{\ifnum\minutes<10 0\the\minutes
            \else\the\minutes\fi}
\def\clock{
\ifnum\hours=0 12:\minute\ AM
\else\ifnum\hours<12 \hour:\minute\ AM
      \else\ifnum\hours=12 12:\minute\ PM
            \else\ifnum\hours>12
                 \pmhours=\hours
                 \advance\pmhours by -12
                 \the\pmhours:\minute\ PM
                 \fi
            \fi
      \fi
\fi
}

\def\monthname{\rel@x\ifcase\month 0/\or January\or February\or
   March\or April\or May\or June\or July\or August\or September\or
   October\or November\or December\else\number\month/\fi}

\def\bold#1{\setbox0=\hbox{$#1$}%
     \kern-.025em\copy0\kern-\wd0
     \kern.05em\copy0\kern-\wd0
     \kern-.025em\raise.0433em\box0 }

\def\lsim{\mathrel{\mathpalette\@versim<}}
\def\gsim{\mathrel{\mathpalette\@versim>}}
\def\@versim#1#2{\vcenter{\offinterlineskip
        \ialign{$\m@th#1\hfil##\hfil$\crcr#2\crcr\sim\crcr } }}

\newcount\@tempcntc
\def\@citex[#1]#2{\if@filesw\immediate\write\@auxout{\string\citation{#2}}\fi
  \@tempcnta\z@\@tempcntb\m@ne\def\@citea{}\@cite{\@for\@citeb:=#2\do
    {\@ifundefined
       {b@\@citeb}{\@citeo\@tempcntb\m@ne\@citea\def\@citea{,}{\bf ?}\@warning
       {Citation `\@citeb' on page \thepage \space undefined}}%
    {\setbox\z@\hbox{\global\@tempcntc0\csname b@\@citeb\endcsname\relax}%
     \ifnum\@tempcntc=\z@ \@citeo\@tempcntb\m@ne
       \@citea\def\@citea{,}\hbox{\csname b@\@citeb\endcsname}%
     \else
      \advance\@tempcntb\@ne
      \ifnum\@tempcntb=\@tempcntc
      \else\advance\@tempcntb\m@ne\@citeo
      \@tempcnta\@tempcntc\@tempcntb\@tempcntc\fi\fi}}\@citeo}{#1}}
\def\@citeo{\ifnum\@tempcnta>\@tempcntb\else\@citea\def\@citea{,}%
  \ifnum\@tempcnta=\@tempcntb\the\@tempcnta\else
   {\advance\@tempcnta\@ne\ifnum\@tempcnta=\@tempcntb \else \def\@citea{--}\fi
    \advance\@tempcnta\m@ne\the\@tempcnta\@citea\the\@tempcntb}\fi\fi}

\catcode`\@=12 

\begin{document}
\vskip 20pt
\def\beq{\begin{equation}}
\def\eeq{\end{equation}}
\def\bea{\begin{eqnarray}}
\def\eea{\end{eqnarray}}
\def\MSbar {\hbox{$\overline{\hbox{MS}}\,$}}
\def\smallMSbar {\hbox{$\overline{\hbox{{\footnotesize MS}}}\,$}}
\def\aBj{\hbox{$a^*_{Bj}$}}
\def\Obs{\hat{O}}
\def\CI{{\cal C}}
\def\naive{na\"{\i}ve}
\def\etal{{\em et al.}}
\newcommand{\mycomm}[1]{\hfill\break{ \tt===$>$ \bf #1}\hfill\break}
\newcommand{\eqref}[1]{eq.~(\ref{#1})}   
\newcommand{\eqrefA}[1]{(\ref{#1})}   
\def\newline{\hfill\break}
\def\tfrac#1#2{{#1/#2}} 
\def\ra{\hbox{$\rightarrow$}\ }

\parskip 0.3cm
\begin{titlepage}
\begin{flushright}
CERN-TH-96/188\\
TAUP-2355-96\\
LBNL-39135\\
hep-ph/9607404
\end{flushright}

\begin{centering}
{\large{\bf
Renormalization-Scheme Dependence of Pad\'e Summation in QCD\\
}}
\vspace{.4cm}
{\bf John Ellis\footnote{The work of J.E. was supported in part
by the Director, Office of Energy
Research, Office of Basic Energy Science of the U.S. Department of
Energy, under Contract DE-AC03-76SF00098.}
}\\
\vspace{.05in}
Theoretical Physics Division, CERN, CH-1211 Geneva 23, Switzerland \\
e-mail: john.ellis@cern.ch \\
\vspace{0.4cm}
{\bf Einan Gardi} and {\bf Marek Karliner}\\
\vspace{.05in}
School of Physics and Astronomy
\\ Raymond and Beverly Sackler Faculty of Exact Sciences
\\ Tel-Aviv University, 69978 Tel-Aviv, Israel
\\ e-mail: gardi@post.tau.ac.il, marek@vm.tau.ac.il
\\
\vspace{0.4cm}
and\\
\vspace{0.4cm}
{\bf Mark A. Samuel}\\
\vspace{.05in}
Department of Physics, Oklahoma State University, \\
Stillwater, Oklahoma 74078, USA\\
e-mail: physmas@mvs.ucc.okstate.edu   \\ 
\vspace{0.4cm}
{\bf Abstract} \\
\bigskip
{\small
We study the renormalization-scheme
(RS) dependence of Pad\'e Approximants (PA's),
and compare them with the Principle of Minimal Sensitivity (PMS)
and
the Effective Charge (ECH) approaches. Although the formulae provided
by the PA, PMS and ECH predictions for higher-order terms in a QCD
perturbation expansion differ in general, their predictions
can be very close numerically for a wide range of renormalization schemes.
Using the Bjorken sum rule as a test case, we find that Pad\'e
Summation (PS) reduces drastically the RS dependence of the
Bjorken effective charge. We use these results to estimate the
theoretical error due to the choice of RS in the extraction of
$\alpha_s$ from the Bjorken sum rule, and use the available
data at $Q^2 = 3$ GeV$^2$ to estimate
$\alpha_s(M_Z) = 0.117^{+0.004}_{-0.007} \, \pm 0.002$, where the
first error is experimental, and the second is theoretical.
} 
\end{centering}
\vfill
\begin{flushleft}
CERN-TH-96/188\\
July 1996 \hfill
\end{flushleft}
\end{titlepage}
\vfill\eject

\section{Introduction}
Pad\'e approximants (PA's) have proven to be useful in many physics
 applications, including condensed-matter problems and
quantum field theory \cite{padeworks}.
PA's may be used either to predict the next term in some
perturbative series, called a Pad\'e Approximant prediction (PAP),
or to estimate the sum of the entire series, called Pad\'e
Summation (PS). The underlying reasons for the successes
of these different applications have not always been apparent. Admittedly,
rational functions are very flexible, and hence {\it
a priori} well suited to approximate other unknown functions,
but some of the PA successes seem almost
`magical'. Obtaining a deeper understanding of these successes is not only
desirable in itself, but may give us deeper understanding also of the underlying
physics. Among the areas in which PA's have had remarkable successes has been
perturbative QCD \cite{SEK,PBB}
where PA's applied to low-order perturbative series have
been shown to `postdict' accurately known higher-order terms, and also used to
make estimates of even higher-order unknown terms that agree with independent
predictions based on the Principle of Minimal Sensitivity (PMS) \cite{PMS}
and Effective Charge (ECH) \cite {ECH} techniques.
Of particular interest to us has been the perturbative QCD series for the
Bjorken sum rule for three quark flavors \cite{BJ,Kodaira,BJcorr}
which has served us previously \cite{PBB}
 as a test case\footnote{Note that in this paper we use PA's for the
effective charge, rather than
for the Bjorken sum rule itself, motivated in part by the large-$N_f$
analysis of \cite{LTM}.}.
For this series, the [0/1] PAP for the third-order
coefficient in the \MSbar\ prescription is $12.8$,
to be compared with the PMS estimate of $20.0$, the ECH estimate of
$19.2$, and the exact value \cite{BJcorr} of $20.21$\ .
This lowest-order PAP is pointing in the right direction, which is
the best one could hope at this level. Going to the next order,
the [1/1] and [0/2] PAP's for the fourth-order Bjorken sum rule coefficient
are $114$ and $99$ respectively,
whilst the PMS/ECH prediction is $130$ \cite{Kataev}.
These values are quite similar, and we have provided \cite{PBB}
a prescription for
systematic improvement of these PAP's which brings them even closer to
the PMS/ECH prediction. Should the previous agreement of PAP's with the
PMS/ECH and
exact calculations here and elsewhere, and the agreement of these
new predictions, be regarded as fortuitous, or is there some deeper
reason why PAP's and PS's should be believed also in the QCD context?

In recent papers \cite{SEK,PBB},
we have tried to cast some light on these `magical'
successes. In particular, we have proven that certain conditions on the
ratios of consecutive terms in a series are mathematically
sufficient to guarantee rapid convergence of successive PAP's, and we have
observed that these conditions are satisfied by asymptotic series dominated
by one or a finite number of renormalon poles. This is believed to be the case
for many QCD perturbation series, such as that for the Bjorken sum rule
\cite{LTM}, which
we have used as a testing ground and showcase. We have also shown that PA's
yield a renormalization-{\em scale} dependence which is much less than that of
the corresponding perturbative series, even when the latter
is supplemented by an ECH estimate
of the next, uncalculated term. Since the full QCD expression for any
physical quantity such as the Bjorken sum rule must be renormalization-scale
independent, this is valuable circumstantial evidence that the PA's are
indeed converging towards the correct physical result.
However, the strength of this evidence is significantly reduced by the fact
that the scale dependence has been studied only within one specific
renormalization {\em scheme}, namely \MSbar, making it difficult
to assess the numerical accuracy of the PA prediction.

The purpose of this paper is to understand better the
renormalization-{\it scheme} dependence of PA's in perturbative QCD,
again taking the
Bjorken sum rule as our test case. On the way to this goal, we also
examine more closely the relations between PAP's and the PMS and ECH
techniques for estimating higher-order
perturbative coefficients in QCD, and examine the extent to which
PS's should and do agree with PMS and ECH estimates of the `sums' of
perturbative series in QCD. We
also examine the extent to which the Cancelation Index (CI) criterion of
Ref.~\cite{Raczka} provides a reliable guide to the comparative accuracies
of partial calculations in different renormalization schemes.

The PMS and ECH formulae used to predict the next term in any QCD
perturbative series do not in general coincide, and we show
below that the PA
prediction (PAP) for the next term is in general
different again. However, in a wide range of RS's their predictions can be
quite close numerically, as we discuss later.

Using  the Bjorken sum rule as an example, we exhibit a map of its
two-parameter renormalization-scheme dependence at the
next-to-next-to-leading order (NNLO) level, situating on this map the PMS
and ECH scheme choices. We then exhibit the corresponding map for the
[0/2] PA, which we find to be much less sensitive to the choice of
renormalization scheme. We also demonstrate that the CI criterion of
Ref.~\cite{Raczka} selects efficiently the region of
renormalization-scheme space where the scheme-dependence is minimized. In
addition, we compare the PMS/ECH and [0/2] PA predictions for the fourth-order
term in the Bjorken sum rule series, finding remarkable agreement.

Finally, as an application of this analysis, we revisit the
extraction \cite{PBB}
of $\alpha_s$ from experimental data on the Bjorken sum
rule at $Q^2 = 3$ GeV$^2$. Our analysis enables us to assign a systematic
error to the choice of renormalization scheme, which is small compared
with the current experimental error. The present data yield
\beq
\alpha_s(M_Z) \, = \, 0.117\,\,{}^{+0.004}_{-0.007}\,\hbox{(exp.)}
 \,\, {}^{+0.002}_{-0.002}\,\hbox{(th.)}
\label{present}
\eeq
which could in the future become a highly competitive determination of
$\alpha_s(M_Z)$, if the present experimental error could be halved.
The systematic error associated with the choice of
renormalization scale would still not be dominant at this level.

\section{Comparing PMS, ECH and Pad\'e Predictions to NNLO}
We start by recalling the essential physical ideas of the
PMS \cite{PMS} and ECH \cite{ECH}
approaches. The PMS method is based on choosing
the RS that minimises the RS dependence in a given order of
perturbation theory, whereas, in the ECH method, one
chooses a natural RS to describe the observable, namely one
in which all the non-leading corrections are exactly zero.
To see how these work and differ in practice, we
look at the application of PMS and ECH to the
perturbative series for a generic
QCD observable, calculated at NLO and
NNLO in some RS. At the NLO level, the choice of RS
involves just the choice of renormalization scale $\mu$,
whereas a second parameter enters at the NNLO
level, as we discuss in more detail later.
For convenience, instead of $\mu $,
we use $\tau $, defined by
\beq
\tau = b \ln \left( \frac {\mu}{ \Lambda } \right)\,,
\qquad\qquad
b=\frac{33-2N_f}{6}\,.
\eeq
At NLO any observable $\Obs$ can be written as:
\beq
\Obs=a(\tau)[1+r_1(\tau)a(\tau)]
\label{Obs_NLO}
\eeq
where $a(\tau) \equiv {\alpha_s(\tau)}/{\pi}$
satisfies the renormalization-group equation at NLO:
\beq
\frac {\partial a}{\partial \tau} = \frac{1}{b} \mu
 \frac {\partial a}{\partial \mu}= -a^2 \left[ 1+ca \right]\,,
\qquad
\qquad
\qquad
c=\frac{153-19N_f}{2(33-2N_f)}\,.
\label{RG_equation}
\eeq
which has the solution:
\beq
\tau = \frac{1}{a} + c \ln \left(\frac{ca}{1+ca}\right)
\label{RG_solution}
\eeq
In the PMS method  \cite{PMS} one chooses an `optimal RS' which
minimises the RS dependence at a given order. The corresponding parameters
are denoted by:
$\overline{\tau}$,$\overline{a}$, $\overline{r_1}$.
In order to find the PMS RS, one differentiates (\ref{Obs_NLO}) with
respect to $\tau $, substituting $\partial a /\partial \tau$ from
(\ref{RG_equation}), yielding
\beq
\frac {\partial \Obs}{\partial \tau} = a^2 \left(-1+\frac{\partial
   r_1}{\partial \tau} \right) - a^3(2r_1 c a + c + 2 r_1)
\label{diff_Obs_NLO}
\eeq
Clearly, at NLO the ${\cal O}(a^2)$ RS dependence of any
observable must vanish, and therefore ${\partial r_1}/{ \partial \tau} =1$.
Thus one identifies the RG invariant $\rho_1$:
\beq
\rho_1 = \tau - r_1
\label{rho_1}
\eeq
which enables us to calculate $r_1(\tau)$ in any RS from an initial
$r_1$ that was calculated in perturbation theory in some initial RS.
Equating ${\partial \Obs}/{\partial \tau} $ in
(\ref{diff_Obs_NLO}) to zero, one finds that in the PMS RS:
\beq
\overline{r_1} = -\,\frac{c}{2(1+c \overline{a})}
\label {r_1_PMS}
\eeq
and therefore, using (\ref{rho_1}), (\ref{RG_solution}),
 and (\ref{r_1_PMS}) one can write an
equation for $\overline{a}$:
\beq
\rho_1 = \frac{ c}{2(1+c \overline{a})} +
\frac{1}{\overline{a}}
 + c \ln \left(\frac{c\overline{a}}{1+c\overline{a}}\right)
\eeq
This equation cannot be solved analytically, but it can directly be
solved numerically.
Finally, one uses (\ref{r_1_PMS}) to write the observable in the PMS RS:
\beq
\Obs_{PMS}=\overline{a} \frac{2+c\overline{a}}{2(1+c\overline{a})}
\eeq
Substituting $\overline{a}$ as a power series in the original $a$,
 we obtain a ratio of two
polynomials in $a$, that is -- a structure similar to Pad\'e-Summation!
With the resemblance comes the difference:
The PMS result depends not only on the lower order coefficients of the
observable we are dealing with, but also on the coefficients of the
$\beta$ function (this is also the case in the BLM method).
\vskip 14pt
In the ECH method \cite{ECH}, the preferred RS is the one is which the
perturbative expansion (like eq~(\ref{Obs_NLO})\ ) reduces to a
leading term, that is:
\beq
\Obs_{ECH} = a^*.
\eeq
We can find this RS by substituting $r_1=0$ in (\ref{rho_1}). Using
(\ref{RG_solution}) we can write the following equation for $a^*$:
\beq
\rho_1 = \frac{1}{a^*} + c \ln \left(\frac{ca^*}{1+ca^*}\right)
\label{NLO_ECH_equation}
\eeq
As in the PMS case, this equation can only be solved numerically. In
this case, we do not find any resemblance to the Pad\'e structure.

At NNLO $a \equiv \alpha_s(\tau)/\pi$ satisfies the RG equation:
\beq
\frac {\partial a}{\partial \tau} = \frac{1}{b} \mu
 \frac {\partial a}{\partial \mu}= -a^2 \left[1+ca+c_2 a^2 \right]
\equiv\frac{\beta{(a)}}{b}
\label{RG_equation_NNLO}
\eeq
which has the formal solution:
\beq
\tau = \frac{1}{a} + c\, \ln \left(\frac{ca}{1+ca}\right)+c_2 \int_0^a
\frac{dx}{(1+cx)\,(1+cx+c_2x^2)}
\label{RG_solution_NNLO}
\eeq
The two independent parameters specifying the RS may be
chosen to be $a$ and $c_2$ or $\tau $ and $c_2$. For our
present purposes, it is more convenient to use $\tau $ and $c_2$.
A generic QCD observable at the NNLO may then be written in
the form:
\beq
\Obs=a\, \left[\,1\,+\,r_1(\tau, c_2)\,a\,+\, r_2(\tau, c_2)\,a^2\,
\right] \,,
\label{Obs_NNLO}
\eeq
where $ a\equiv a(\tau, c_2)$.
To derive the second-order PMS
formulae, we first
differentiate the observable (\ref{Obs_NNLO}) with respect to both
$\tau $ and $c_2$:
\beq
\frac {\partial \Obs}{\partial \tau} = \frac {\partial a}{\partial \tau}
\left(1+2r_1a+3r_2a^2\right)+ a^2 \,\frac {\partial r_1}{\partial \tau}
+a^3 \,\frac {\partial r_2}{\partial \tau}
\label{obs_diff_tau}
\eeq
\beq
\frac {\partial \Obs}{\partial c_2} =\frac {\partial a}{\partial c_2}
\left(1+2 r_1 a + 3 r_2 a^2 \right)
+ a^2 \,\frac {\partial r_1}{\partial c_2}
+a^3 \, \frac {\partial r_2}{\partial c_2}
\label{obs_diff_c_2}
\eeq
Using next the NNLO renormalization-group equation
(\ref{RG_equation_NNLO}), we can find
 $\tfrac {\partial a}{\partial c_2}$
 as a function of $a$, $c$ and $c_2$ and substitute
 $\tfrac {\partial a}{\partial \tau}$ and
 $\tfrac {\partial a}{\partial c_2}$ for the appropriate power series in $a$
in equations (\ref{obs_diff_tau}) and (\ref{obs_diff_c_2}).
Demanding that
 ${\partial \Obs}/{\partial \tau} $ and  ${\partial \Obs}/{\partial c_2} $
 be of order $O(a^4)$,
we find two renormalization-group-invariant quantities:
\beq
\rho_1 = \tau - r_1
\label{rho_1_again}
\eeq
which appears already in a NLO analysis, and
\beq
\rho_2 = r_2 +c_2 - r_1^2-r_1c
\label{rho_2}
\eeq
Using these two invariants, we can calculate $r_1(\tau,c_2)$ and
$r_2(\tau,c_2)$ in any RS, as a function of the values of
$r_1$ and $r_2$ calculated in perturbation theory in
any initial RS. We will use (\ref{rho_1_again}) and
(\ref{rho_2}) extensively later in this work.
\vskip 14pt
The two equations that determine the PMS RS can now be found by
equating (\ref{obs_diff_tau}) and (\ref{obs_diff_c_2}) to zero.
These equations cannot be solved analytically,
but one may solve them graphically to locate
the PMS RS,
by plotting the NNLO observable as a function of the RS
parameters $a$ and $c_2$,
and identifying a local extremum or saddle point, which
corresponds to both (\ref{obs_diff_tau}) and (\ref{obs_diff_c_2})
being zero.
\vskip 14pt
The ECH RS is specified at third order by the conditions
$r_1=r_2=0$, which we must substitute
into equations (\ref{rho_1_again}) and (\ref{rho_2}). The
results are:
\beq
\tau_{\, ECH} = \rho_1
\eeq
and,
\beq
c_{2\, ECH} = \rho_2
\eeq
Substituting the above in (\ref{RG_solution_NNLO}), we obtain an
equation for $a^*$, which is just the observable calculated in the
ECH scheme, $\Obs_{ECH} = a^*$:
\beq
\rho_1 = \frac{1}{a^*} + c \ln
\left(\frac{ca^*}{1+ca^*}\right)+\rho_2 \int_0^{a^*}
\frac{dx}{(1+cx)(1+cx+ \rho_2  x^2)}
\label{NNLO_ECH_equation}
\eeq
As in the PMS case, this ECH equation cannot be solved analytically.

 \vskip 14pt
Since there are no closed analytical formulae for the third-order
PMS and ECH results, we cannot compare them directly to the PS
method. We do note, however, that the PMS and ECH expressions
contain in general singularities in the coupling plane,
as do Pad\'e approximants. Still, the
PMS singularities are not necessarily discrete poles, and thus the PMS
and ECH singularity structure may differ from that of the PS.
Here we do not address this interesting issue further, focusing instead
on comparing Pad\'e with PMS/ECH. The comparison is done
by looking both at their
predictions of for the next term in the perturbative
series, and at their results for the overall `summation' of the series.

As a warm-up exercise, we consider the comparison at NLO, where we
start with (\ref{Obs_NLO}) and predict $r_2$ by the different methods.
To this order, the [0/1] PAP is equivalent to the assumption that
the series is a geometrical one, so that:
\beq
r_2^{PAP}=r_1^2
\label{NLO_PAP_prediction}
\eeq
On the other hand, it follows from the above
higher-order analysis that the PMS prediction is~\cite{PMS}:
\beq
r_2^{PMS}=\left( r_1+\frac{1}{2}c \right)^2
\label{NLO_PMS_prediction}
\eeq
whilst the ECH result is~\cite{ECH}:
\beq
r_2^{ECH}=r_1^2+r_1 c
\label{NLO_ECH_prediction}
\eeq
We observe that, if $r_1$ is much larger than $c$, then the three
methods will give similar results. However, $c$ does not
depend on the RS, while $r_1$ does. This means that
there are schemes
in which the PAP, PMS and ECH would be close, and
others (which may be just as legitimate) in which they would not
agree. It is however clear from equations (\ref{NLO_PMS_prediction})
and (\ref{NLO_ECH_prediction}) that the schemes in which there is a good
agreement between the PMS and ECH predictions
are {\it exactly} those schemes in which the PAP prediction
(\ref{NLO_PAP_prediction}) agrees with both of them.
\vskip 14pt
In order to study this comparison between Pad\'e and PMS further,
we go to third order, where a generic observable may be
written as in (\ref{Obs_NNLO}).
The PAP predictions for the 4-th coefficient ($r_3$) are:
\beq
r_3^{[1/1] \, PAP}=\frac{r_2^2}{r_1}
\label{r311pap}
\eeq
\beq
r_3^{[0/2] \, PAP}=-r_1^3+2r_1r_2
\label{r302pap}
\eeq
whilst the PMS and ECH predictions coincide exactly \cite{Kataev}:
\beq
r_3^{PMS}=r_3^{ECH}=r_1 \left( 3r_2-2r_1^2+c_2-\frac{cr_1}{2} \right)
\label{r3pmsech}
\eeq
We see from the above that
there is no simple relation between the
formulae for the PAP and the PMS/ECH
predictions.

At NLO, eqs.~(\ref{NLO_PAP_prediction}),
(\ref{NLO_PMS_prediction}), (\ref{NLO_ECH_prediction}),
they differ only by terms that
depend on a higher-order coefficient in the QCD $\beta$ function.
However, there are further differences
at NNLO, eqs.~(\ref{r311pap}), (\ref{r302pap}), (\ref{r3pmsech}).
There are, nevertheless, a couple of
extreme cases in which the predictions of the different
  methods for $r_2$ and for $r_3$ approach each other.
One such case is provided by
  RS's in which all the non-leading corrections are small,
 or even zero as in the ECH RS.
  The prediction for the next term is then small in all approaches,
  as intuitively expected.
  The other
  extreme is when the non-leading corrections are
  much larger than the corresponding coefficients of the $\beta$ function
  ($r_i \gg c_i$)\footnote{An example of this class
seems to be provided by the Bjorken sum rule in
\MSbar\ with $\mu=Q$.}.
In this case the NLO prediction is $r_2^{est} \sim r_1^2$ in any
method, and the NNLO prediction is $r_3^{est} \sim r_1^3$ in any method,
provided that the $r_2$
coefficient is indeed close to the NLO prediction, i.e.,
$r_2 \sim r_1^2$.
It cannot be guaranteed, however, that either of these conditions
holds for a generic observable in a generic RS.
Therefore, the formulae for the next term derived
in the different methods do not generally
coincide. Still,
a good {\em numerical} agreement between the predictions is possible
in a generic RS, and this is indeed the case in
the example we discuss later.

The principal conclusions of this analysis are:
\begin{description}
\vskip 10pt
\item{a) \,} There is no general agreement, nor
simple relation between the PMS or the ECH methods and the
Pad\'e method. One major reason for this is that, in contrast to the
PA's, the PMS and ECH predictions depend on the
coefficients of the $\beta$ function, and
not only on the coefficients of the observable under
consideration.
\vskip 10pt
\item{b) \,} PA's differ from conventional perturbation series by
having distinctive
singularities in the coupling plane. While some singularities may be
expected in QCD, they are not necessarily of the Pad\'e form.
Singularities in the
coupling plane appear also in the PMS and ECH formulations, e.g.,
the second-order PMS result is just a rational
polynomial.
 Higher-order PMS and ECH results cannot be calculated
analytically, but singularities in the
coupling plane are expected there as well. However, there is at
present no indication that the
singularity structure will resemble that of the PA.
\vskip 10pt
\item{c) \,} Since the PA method is totally
independent of the PMS and ECH methods,
we believe that good numerical agreement
between their predictions should
be considered as strong evidence that
both sets of predictions are correct.
\end{description}
\vskip 14pt
As a test case for the application of the PMS, ECH and PA
approaches, we study in the following sections
the Bjorken Sum Rule for three
quark flavours. Our conclusions can later be checked using
other QCD observables.
\vskip 14pt
\section{Renormalization-Scheme Dependence in the \newline Bjorken Sum Rule}

We now proceed to a detailed discussion of renormalization-scheme
(RS) dependence in one of the cases where an exact NNLO perturbative
calculation in available, namely the effective charge in the
Bjorken sum rule \cite{BJ,Kodaira,BJcorr}:
\beq
a^*_{Bj} = a \,+\, 3.58 \, a^2 \,+\, 20.22 \, a^3 \,+ \,\cdots
\label{Bj}
\eeq
in the \MSbar\ scheme with $\mu=Q$ and 3 flavours assumed.
It is a simple matter to convert \eqrefA{Bj} to an arbitrary
renormalization scheme, using the two RG invariants
of eq.~(\ref{rho_1_again}) and (\ref{rho_2}).

We note at the outset of our analysis that the PMS and ECH
prescriptions are both based explicitly on the
renormalization group, and aim directly at the choice
of an optimal RS. The BLM approach \cite{BLM}
is also based on the
renormalization group, and can be regarded as a way of
choosing systematically renormalization scales
appropriate for each order in perturbation theory.
As we discuss in \cite{RS_BLM}, the BLM approach is close to
PMS and ECH, both in its nature and its results\footnote{There
are some intriguing connections between the
  mathematical foundations of the BLM and Pad\'e approaches. These are
  currently under investigation \cite{RS_BLM}.}.
On the other hand, Pad\'e approximants are formulated independently
of the renormalization group in any RS, and we have demonstrated
explicitly in the previous section that they are not related
to PMS and ECH in any obvious way.
Hence, there is no {\em a priori} reason to expect the Pad\'e
method to reduce the RS dependence. In fact it does,  as we
shall see later, and we believe that this observation bolsters
the utility of Pad\'e approximants in QCD applications\footnote{We
demonstrated previously that the Pad\'e method
greatly diminishes the renormalization {\em scale} dependence of the
Bjorken sum rule.}.

We first compute the Bjorken effective charge \aBj\ as
a function of the two NNLO parameters $a$
and $c_2$ discussed in the previous section.
Experimental measurements of the Bjorken sum rule are currently made
in a range of $Q^2$ where one believes 3 quark flavours to be
active, as assumed in \eqrefA{Bj}. In principle, as
$\mu$ is varied, one may cross the charm threshold, and so one should
modify and match formulae \eqrefA{RG_equation_NNLO}
and  \eqrefA{Bj} of effective
theories at the $N_f=4$ threshold. Since this issue is only a technical
complication, we choose to avoid it for the purposes of this discussion
by calculating the Bjorken effective charge at
$Q^2 = 20$ GeV$^2$, corresponding to $a=0.07$ in the \MSbar\
prescription with $\mu=Q$, and fixing $N_f=3$, whatever the
value of $\mu$. This analysis is sufficient to establish
a ``proof of concept", and we return later to a discussion of the
more experimentally relevant case of lower $Q^2$.
Fig.~1 displays contours of the Bjorken effective charge
\aBj, differing in height by $\Delta\aBj=0.002$, where we note
the following features:
\begin{description}
\vskip 10pt
\item{a) \,} There is a flat region
  around $a=0.1$, where the RS dependence is very weak.
\item{b) \,} The value of the Bjorken effective charge in the \MSbar\
  RS ($a=0.07$, $c_2=4.471$, denoted by a circle) is
\beq
\aBj(\smallMSbar)=0.09449
\label{3rd_BjSR_result}
\eeq
We note that the \MSbar\ RS does not lie in the flat region
mentioned above. Therefore the RS dependence, particularly
the renormalization scale dependence, is relatively high in the \MSbar\ RS.
\vskip 10pt
\item{c) \,} Within the flat region mentioned in a),
there is a saddle point at $a=0.1005$, $c_2=8.7$, which  corresponds to
the PMS RS \footnote{The exact location of this saddle point was found
using a similar plot with much higher resolution, which is not
presented here.}. The value of the  Bjorken effective charge in
the PMS RS is
\beq
\aBj(\hbox{\footnotesize PMS})=0.10033
\label{PMS_result}
\eeq
which deviates by about $ 6 \% $ from the $ \MSbar $ result. This
deviation is an example of the RS dependence in this case.
\vskip 10pt
\item{d) \,} Another point which lies in the flat
  region mentioned in a) is the ECH RS at $a=0.10038$,
  $c_2=5.476$. The value of the Bjorken effective charge in
  the ECH method is therefore
\beq
\aBj(\hbox{\footnotesize ECH})=0.10038
\label{ECH_result}
\eeq
which is very close to the PMS result of (\ref{PMS_result}).
\vskip 10pt
\item{e) \,} When using a RS with a very low coupling-constant
 ($a \lsim 0.04$, for example)
  or a very high coupling-constant ($a \gsim 0.16$, for example)
we obtain a value of \aBj\ which is
totally inconsistent with the \MSbar, PMS and ECH results,
and which is also
strongly dependent on the specific choice of the parameters $a$
and $c_2$. This strong deviation from the results of
the ECH and PMS RS's is related to the existence of
large non-leading corrections
in the perturbative expansion for the Bjorken Sum Rule
in these RS's. Therefore, we look for a consistent way of
excluding these RS's, or -- even better -- a consistent way of
using them and still getting reasonable results. We will now
show that both aims are achievable,  the first by the use of the
Cancellation Index criterion advocated in Ref.~\cite{Raczka}, which is
discussed in the following section, and the second  by the use
of PS, as shown in section 5.
\end{description}

\vskip 14pt
\section{The Cancellation Index Criterion}
\vskip 14pt

In view of the NNLO RS dependence on $a$ and $c_2$ displayed
in Fig.~1, it is desirable to find a criterion which selects
a region in the $(a,\,c_2)$ plane that contains
``well-behaved" RS's for which higher-order corrections are not
expected to be large. One can then examine the performance
of techniques for improving the perturbative series (such as
the Pad\'e method) in a compact domain of the
$(a,\,c_2)$ plane. Looking at the RS dependence over this domain
may then provide a legitimate estimate of the observable
(\aBj\ in our test case) and of the RS uncertainty in this estimate.

Here we use the criterion proposed in Ref.~\cite{Raczka},
namely that a ``well-behaved" RS is one for which the degree of
cancellation between the different terms in the second NNLO
renormalization-group invariant
$ \rho_2 = r_2 + c_2 - r_1^2 - c r_1 $
\eqrefA{rho_2} is small. The degree
of cancellation is measured by the Cancellation Index:
\begin{equation}
\label{CI_criterion}
\CI =
\frac{|r_2| + |c_2| + r_1^2 + c |r_1|}{r_2 + c_2 - r_1^2 - c r_1}
\end{equation}
and contours of $\CI$\ for the Bjorken effective charge \aBj\
are also plotted in Fig.~1\,. We exhibit the contours
$\CI=2,3,4,5$: contours of higher values of $\CI$\ become closer
together as $\CI$\ increase.

We observe that these contours of $\CI$\ are indeed centered around
the flat region of small RS dependence to which we drew attention
previously. Indeed, the ECH RS is the only one for which
$\CI=1$. The PMS RS also has a low value $\CI=2.18$, whereas
$ \CI \gsim 7 $ for the \MSbar\ RS, as was already mentioned in
\cite{Raczka}.

In order to study the RS dependence we restrict
our attention to the domain defined by $\CI \leq \CI_0$. $\CI_0$
should be chosen such that the PMS RS, where the local RS dependence
vanishes is well within the selected domain, but yet, not
too large, so that all the RS included in the domain would be
trustable, having a small enough local RS dependence.
For the purposes of the subsequent discussion, we shall choose
 $\CI_0=4$\, which answers the above requirements.
 While other reasonable values of $\CI_0$ and other criteria for
 choosing restricted domains in the RS parameter space may be used just
 as well, our principle conclusions would remain the same.

Within the $\CI \leq 4$ domain, we find
\beq
0.087 \, \leq \, \aBj \, \leq \, 0.109
\label{aBj_range}
\eeq
which is our best estimate of the likely RS ambiguity
in \aBj, in the absence of the improvement that the Pad\'e
method provides in the next section.
\vskip 14pt
\section{Pad\'e Summation of the Bjorken Series}
\vskip 14pt

We now apply the Pad\'e method to the perturbative QCD series
for the Bjorken effective charge \eqrefA{Bj}, and explore its
effect on the RS dependence.
As already mentioned in the Introduction, one may use
Pad\'e approximants either to predict the next term in the
series (PAP) or to sum the entire series (PS), in the sense \cite{PBB}
of calculating the Cauchy principal value of an asymptotic
series with one or more infrared renormalons, as is believed
to be the case for the Bjorken series.

{\em A priori}, one may evaluate either the [0/2] PS or the
[1/1] PS of the Bjorken series. In this section we evaluate
both PS's of the Bjorken series in the $(a,\,c_2)$ plane
introduced earlier, and compare their RS dependences with that of
the \naive\ series shown in Fig.~1.

Fig.~2 displays contours of \aBj\ evaluated using the
[0/2] PS, with the same steps $\Delta\aBj=0.002$ as in Fig.~1\,.
We see immediately that the contours in Fig.~2 are much sparser
than those in Fig.~1, and hence that the [0/2] PS depends much less on the
RS than does the \naive\ partial sum. To put this comparison
on a quantitive basis, we evaluate the RS dependence
of \aBj(\,[0/2]\,PS\,) in the $\CI \leq 4$ domain of the
$(a,\,c_2)$ plane:
\beq
0.0986 \, \leq \, \aBj(\,[0/2]\,PS\,) \, \leq \, 0.1011
\label{02PS}
\eeq
We see that the RS dependence of the [0/2] PS is an order
of magnitude less than that of the \naive\ partial sum
\eqrefA{Bj}.

This is not true, however, for the [1/1] PS shown in Fig.~3, whose
RS-dependence is much larger to that of the \naive\ partial sum:
\beq
-0.14 \, \lsim \,\aBj(\,[1/1]\,PS\,)  \, \lsim \, 0.14
\label{11PS}
\eeq
The extremes of this large range are due to specific RS's for
which the [1/1] PS is particularly deviant. Most RS's fall within
a much narrower range. However, this analysis points up the fact
that the [1/1] PS is much less well-behaved than the [0/2] PS
\eqrefA{02PS}.

A persistent problem in the application of Pad\'e methods is how to choose
the one which is the most accurate. Various empirical and analytic
results give some indications, but there is no unambiguous general
prescription for the choice. In the case of the Bjorken series at the
NNLO level,
the amount of RS dependence provides a clear physical criterion, which
selects unambiguously the [0/2] PS.
The reduced RS dependence \eqrefA{02PS} of the
\aBj(\,[0/2]\,PS\,) hints that this determination of \aBj\ may be
correct within its errors. One might worry that the PS for
different RS are converging to the wrong common result.
However,
encouragement is provided by the comparison between the
PMS, ECH and PS results \eqrefA{PMS_result}, \eqrefA{ECH_result} and
\eqrefA{02PS}, where we see that they are all consistent.

Since the PMS/ECH and the PS methods are {\em a priori} unrelated
-- PMS and ECH are based on the renormalization group and attempt
to minimize higher-order terms, whereas PS uses no
renormalization-group ingredients and tries to resum rapidly-growing
higher-order terms -- we regard the remarkable agreement between
these different techniques as strong evidence in favour of both
methods. Further support for this conjecture comes in the next section,
where we compare PAP's with PMS/ECH predictions for the next term
in the Bjorken series.

\vskip 14pt
\section{PAP and PMS/ECH Predictions for the Next Term in the Bjorken Series}
\vskip 14pt
If the agreement between PS and PMS/ECH is significant, and not just
a coincidence, we can formulate several expectations concerning the
coefficient of the fourth-order term predicted by the different methods:
\begin{description}
\vskip 10pt
\item{a) \,} The fourth-order partial sum with the
 fourth-order coefficient given by the [0/2] PAP
should be consistent with the PS  result of
 \eqrefA{02PS} (and thus also with the PMS and ECH result of
 (\ref {PMS_result}) and (\ref {ECH_result}) ).
It should also have a RS dependence which is smaller
  than the original third order partial sum,
  but larger than the full PS result.
\vskip 10pt
\item{b) \,} The same should hold for the fourth-order partial sum when
  the PMS/ECH prediction for the next term is used for the fourth-order
coefficient.
\vskip 10pt
\item{c) \,} There should be a numerical agreement between the
predictions of the [0/2] PAP and the PMS/ECH
for the fourth-order term in every RS.
\end{description}
\vskip 14pt
We now examine the [0/2] PAP and the PMS/ECH predictions for the
fourth-order coefficient in the Bjorken series, and verify that these
expectations are indeed realized.
Figures 4 and 5 display in the $(a, c_2)$ plane the fourth-order partial
sum of the perturbative QCD series for the Bjorken effective charge
\beq
a^*_{B_j}({\rm 4th-order}) =
 a\left[1+r_1(a,c_2)a+r_2(a,c_2)a^2+r_3^{est}(a,c_2)a^3\right]
\label{NNNLO_obs}
\eeq
where, in Fig. 4, we have used the [0/2] PAP: $r^{est}_3 =
r_3^{[0/2]PAP}$, and in Fig. 5 the common PMS/ECH prediction:
$r_3^{est} = r_3^{PMS} = r_3^{ECH}$. We have included the
$\CI = 4$ contour in both figures.
\vskip 14pt
Comparing Figs. 4 and 5 with 1 and 2, we
see the following.
\begin{description}
\item {1. \,} We see from Fig. 4 that the fourth-order effective charge
 (\ref{NNNLO_obs}),
evaluated with the  [0/2] PAP coefficient within the $\CI < 4$ region, is
\beq
0.098  \, \leq \, a^*_{B_j}([0/2]~PAP) \, \leq \, 0.110
\label{PAP_result_uncertainty}
\eeq
This is consistent with the full PS result (\ref{02PS}) and the PMS and ECH
results (\ref {PMS_result}) and (\ref {ECH_result}).
The RS dependence of this result is much smaller than
that of the third-order partial sum (\ref{aBj_range}),
but much larger than that of the
full PS (\ref{02PS}), in agreement with expectation (a) above.
\item {2. \,} We see from Fig. 5 that the fourth-order effective
  charge, evaluated
with the PMS/ECH prediction for the coefficient within the $C < 4$
region, is
\beq
0.095 \, \leq \, a^*_{B_j}(PMS/ECH) \, \leq \, 0.102
\label{PMS_result_uncertainty}
\eeq
Again, this result is consistent with the full PS result (\ref{02PS})
and the PMS
and ECH results (\ref {PMS_result}) and (\ref {ECH_result}).
The RS dependence of this result is also much
smaller than that of the third-order partial sum (\ref{aBj_range}),
but considerably
larger than that of the full PS (\ref{02PS}), in agreement with expectation (b)
above.
\end{description}
\vskip 14pt
The last point is the direct numerical comparison of the fourth-order terms
predicted by the [0/2] Pad\'e approximant and the PMS/ECH methods. Figure 6
displays $r_3^{est}(a, c_2)a^3$ as calculated by the different
methods, where we see excellent agreement between the [0/2] PAP and the
PMS/ECH predictions for $r_3$ in every RS\footnote{We have also
developed \cite{SEGK} a procedure for improving PAP's by taking into
account the expected asymptotic behavior of the perturbative coefficients,
which may be applied in particular to the [1/1] PAP.}.
\vskip 14pt

\vskip 14pt
\section{Application to the Extraction of $\alpha_s(M_Z)$}

To illustrate the approach described above, we now
use the Bjorken sum rule~\cite{BJ,Kodaira,BJcorr} to
extract a value of $\alpha_s(M_Z)$
\cite{BjSRalphas},\cite{PBB}
from the available polarized deep inelastic scattering
data at $Q^2=3$ GeV$^2$ \cite{E142}-\cite{HERMES}. We do
not attempt to re-evaluate the values these experiments quote
for the integrals $\Gamma_1^{p,n}$, nor their quoted errors
due, for example,
to the extrapolations to $x_{Bj} = 0,1$ or the modeling of
the possible $Q^2$-dependence in $g_1^{p,n}(x_{Bj}, Q^2)$.
Data are also available
at higher values of $Q^2$ \cite{PSFdata}, but
modeling the evolution to $Q^2=3$ GeV$^2$
would introduce an additional systematic error which we
prefer to avoid.
For our illustrative purpose, it is sufficient to use
the 3 GeV$^2$ data alone.
The $Q^2=3$ GeV$^2$ data set we use is as follows:
\vbox{
\def\myquad{\hskip10em}
\bea
\Gamma_1^n&=&{-}0.033 \,\pm\,
0.006\hbox{\,(stat.)}\,\pm\,0.009\hbox{(sys.)}
\myquad\protect\cite{E142update}\nonumber\\
\Gamma_1^n&=&{-}0.032\,\pm\,0.013\hbox{\,(stat.)}
\,\pm\,0.017\hbox{(sys.)} \myquad\protect\cite{HERMES}\nonumber\\
\Gamma_1^p&=&\phantom{-}
0.127\,\pm\,0.004\hbox{\,(stat.)}\,\pm\,0.009\hbox{(sys.)}
\myquad\protect\cite{E143p}\label{DISdata}\\
\Gamma_1^d&=&\phantom{-}
0.042\,\pm\,0.004\hbox{\,(stat.)}\,\pm\,0.009\hbox{(sys.)}
\myquad\protect\cite{E143d}\nonumber
\eea
which may be combined to yield the following combined result for
the Bjorken sum rule
\beq
\Gamma_1^p(3\hbox{GeV}^2) -\Gamma_1^n(3\hbox{GeV}^2) =
0.160 \pm 0.014
\label{expBjSR}
\eeq
}
to be compared with the theoretical calculations of
both perturbative and nonperturbative (higher-twist) effects.
We estimate the latter using \cite{HTrefs}:
\beq
\delta (\Gamma_1^p -\Gamma_1^n) =
{-0.02 \pm 0.01 \over Q^2}
\label{HT}
\eeq
As the basis for our extraction of $\alpha_s$,
we use the ECH RS, in which the [0/2] PS coincides
with the partial sum. We then translate the result to the $\MSbar$
RS with $\mu = Q$, so it can be easily compared with other calculations.
This provides the following estimate of $\alpha_s$ in
the $\MSbar$ RS at $Q^2=3$ GeV$^2$:
\beq
a \equiv {\alpha_s(3 \hbox{GeV}^2)\over\pi} = 0.102^{+0.010}_{-0.017}
\label{alphas_at_3}
\eeq
where in the central value
we have included the shift due to the higher-twist estimate
(\ref{HT}).
The error quoted in (\ref{alphas_at_3})
is purely experimental, being obtained
directly from our evaluation \eqrefA{expBjSR}. This must be
combined with theoretical error in the higher-twist estimate (\ref{HT}),
$\delta_a(HT) = \pm 0.003$, and
the theoretical error estimated from the minimum and maximum
values of the [0/2] Pad\'e in the $\CI \leq 4$ region, which yields
$\delta_a(RS)= \pm 0.004$.
Thus we find
\beq
a \equiv {\alpha_s(3 \hbox{GeV}^2)\over\pi} = 0.102^{+0.010}_{-0.017}
\,\,{}^{+0.005}_{-0.005}
\label{alphas_at_3_B}
\eeq
where the first set of errors is experimental and the second theoretical.
Finally, evolving (\ref{alphas_at_3_B}) to $M_Z$, we obtain,
\beq
\alpha_s(M_Z) = 0.117 ^{+0.004}_{-0.007} \,\,{}^{+0.002}_{-0.002}
\label{asatmz}
\eeq
where the extrapolation error is negligible, as discussed in
\cite{PBB}. The dominant source of the theoretical error is
$\delta_a(RS)$, with a somewhat smaller contribution from $\delta_a(HT)$.
It is interesting to compare the central value in
(\ref{asatmz}) with what one would obtain as
the \naive\ result in the \MSbar\ RS:
$\alpha_s(M_Z) = 0.123$ , which is outside the theoretical error
range quoted in (\ref{asatmz}).

\section{Conclusions}

We have explored in this paper the relationship between Pad\'e
Approximants and the PMS and ECH techniques for estimating
higher-order coefficients in perturbative QCD. The similarities
between numerical results at the NNLO level may not be coincidences
in certain choices of RS, as we have discussed above.

Pad\'e summation (PS) has the remarkable
property of reducing drastically the RS dependence of the perturbative
series for the Bjorken sum rule with three quark flavors,
if one chooses the appropriate [0/2] Pad\'e. This observation favors
the hypothesis that PS indeed leads us
rapidly to the correct `sum' of the perturbative QCD series. We have
noted also that the Pad\'e Approximant Prediction (PAP) for the next term
in the series is quite successful, although it does not reduce the RS
dependence as much as the [0/2] PS.

We believe that these results support the suggestion that Pad\'e
Approximants may be useful in applications to perturbative QCD, just as
they have proved to be useful in applications to condensed-matter problems
and elsewhere in quantum field theory. As an illustration how the PS
technique may be useful in QCD, we have applied it to the perturbative
series for the Bjorken sum rule, and used it to reduce the theoretical
error associated with the choice of Renormalization Scheme (RS). Present
data at $Q^2 = 3$ GeV$^2$ yield the evaluation (\ref{asatmz}), in which
the theoretical error (given second) is considerably smaller than the
experimental error (given first). This result indicates that the PS
technique may enable a highly competitive value of $\alpha_s(M_Z)$ to
be extracted from future polarized lepton scattering data.

\bigskip
\begin{flushleft}
{\bf Acknowledgements}
\end{flushleft}
We thank A.B.~Zamolodchikov for discussion on the asymptotic behaviour
of the $\beta$ function.
The work of J.E. was supported in part by the Director, Office of Energy
Research, Office of Basic Energy Science of the U.S. Department of
Energy, under Contract DE-AC03-76SF00098.
The research of E.G. and M.K.
was supported in part by the Israel Science Foundation
administered by the Israel Academy of Sciences and Humanities,
and by
a Grant from the G.I.F., the
German-Israeli Foundation for Scientific Research and
Development.

\vskip 14pt

\def\etal{{\em et al.}}
\def\PL{{\em Phys. Lett.\ }}
\def\NP{{\em Nucl. Phys.\ }}
\def\PR{{\em Phys. Rev.\ }}
\def\PRL{{\em Phys. Rev. Lett.\ }}

\vfill\eject
\begin{figure}[htb]
\begin{center}
\mbox{\kern-2cm
\epsfig{file=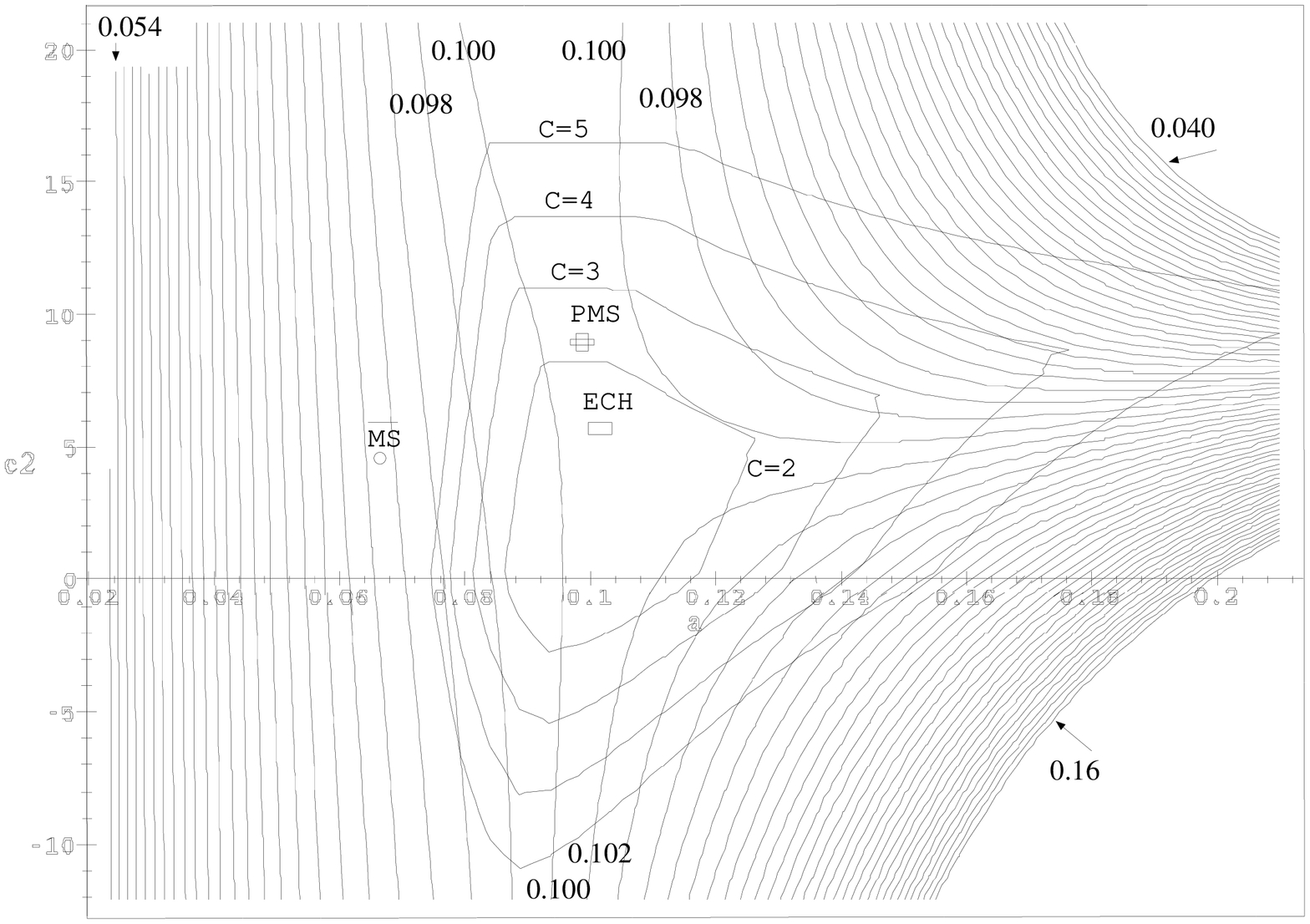,width=17.0truecm,angle=90}}
\caption{
A contour plot in the plane of RS parameters $a, c_2$ of the
Bjorken Sum Rule effective charge $a^*_{Bj}$ calculated up to NNLO, i.e.,
the third-order partial sum. The separations between the contours are
$\Delta a^*_{Bj}=0.002$. The values of $a, c_2$ in the $\MSbar$, PMS and
ECH RS are indicated.  In addition, we plot contours of the Cancellation
Index: $\CI=2,3,4,5$.}
\label{3rd_par_sum}
\end{center}
\end{figure}

\begin{figure}[htb]
  \begin{center}
\mbox{\kern-2cm
\epsfig{file=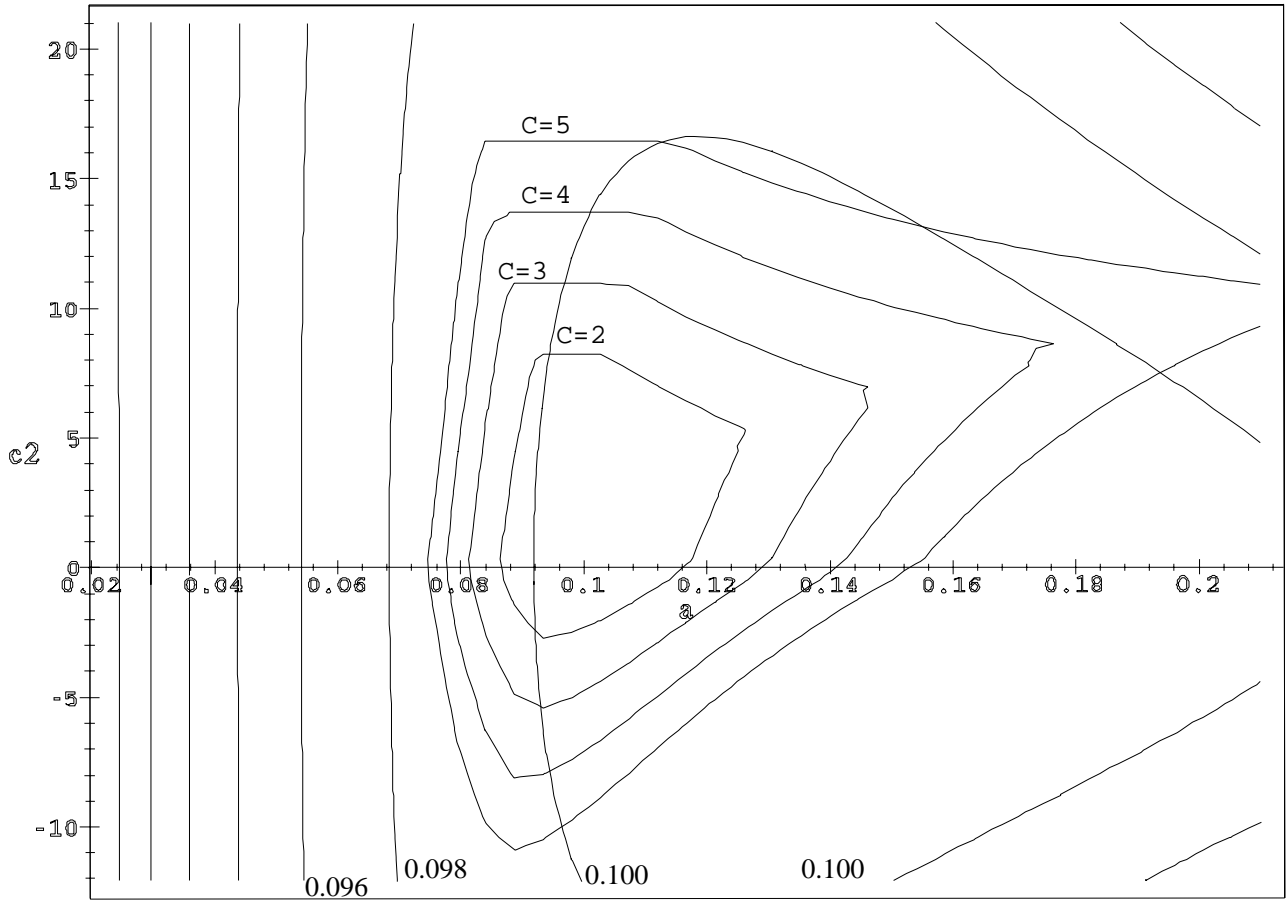,width=17.0truecm,angle=90}}
    \caption{
A contour plot of the RS dependence of the Bjorken Sum Rule
effective charge $a^*_{Bj}$ calculated using [0/2] Pad\'e Summation. The
$a^*_{Bj}$ and $\CI$ contours are spaced as in Fig.~1. The larger
separations between the $a^*_{Bj}$ contours reflect the reduced RS
dependence compared with the third-order partial sum shown in Fig.~1.}
\label{Fig_PS02}
  \end{center}
\end{figure}

\begin{figure}[htb]
  \begin{center}
\mbox{\kern-2cm
\epsfig{file=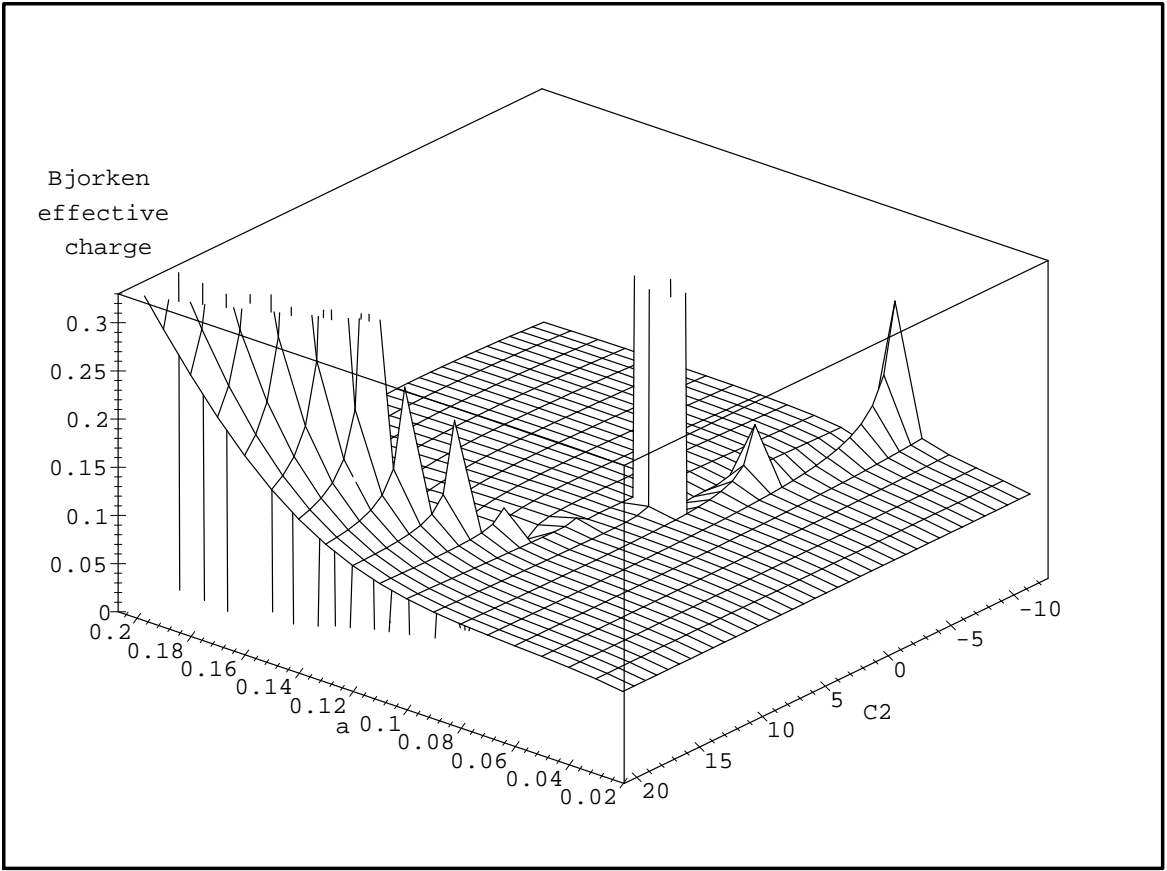,width=17.0truecm,angle=-90}}
    \caption{
The RS dependence of the Bjorken Sum Rule effective charge
$a^*_{Bj}$ calculated using [1/1] Pad\'e Summation. We see in this
three-dimensional plot peaks corresponding to RS's for which the [1/1] PS
is particularly erratic, signalling its unreliability.}
    \label{Fig_PS11}
  \end{center}
\end{figure}

\begin{figure}[htb]
  \begin{center}
\mbox{\kern-2cm
\epsfig{file=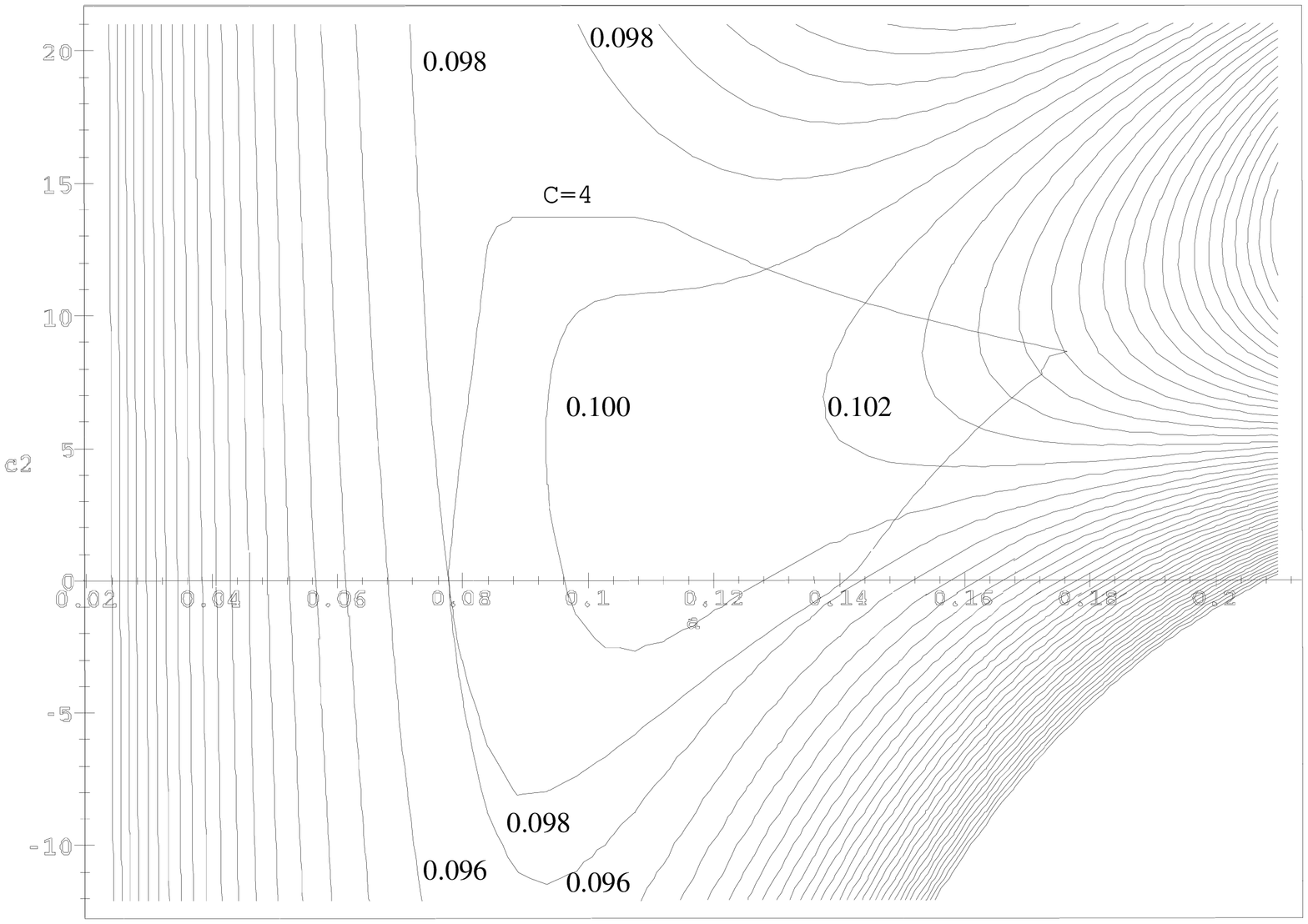,width=17.0truecm,angle=90}}
    \caption{
A contour plot of the RS dependence of the Bjorken Sum Rule
effective charge $a^*_{Bj}$ calculated using a fourth-order partial sum
evaluated with the [0/2] PAP fourth-order coefficient. The $a^*_{Bj}$
contours are spaced as in Fig.~1, and we also show the $\CI=4$ contour.}
    \label{PAP improvement}
  \end{center}
\end{figure}
\vskip 14pt

\begin{figure}[htb]
  \begin{center}
\mbox{\kern-2cm
\epsfig{file=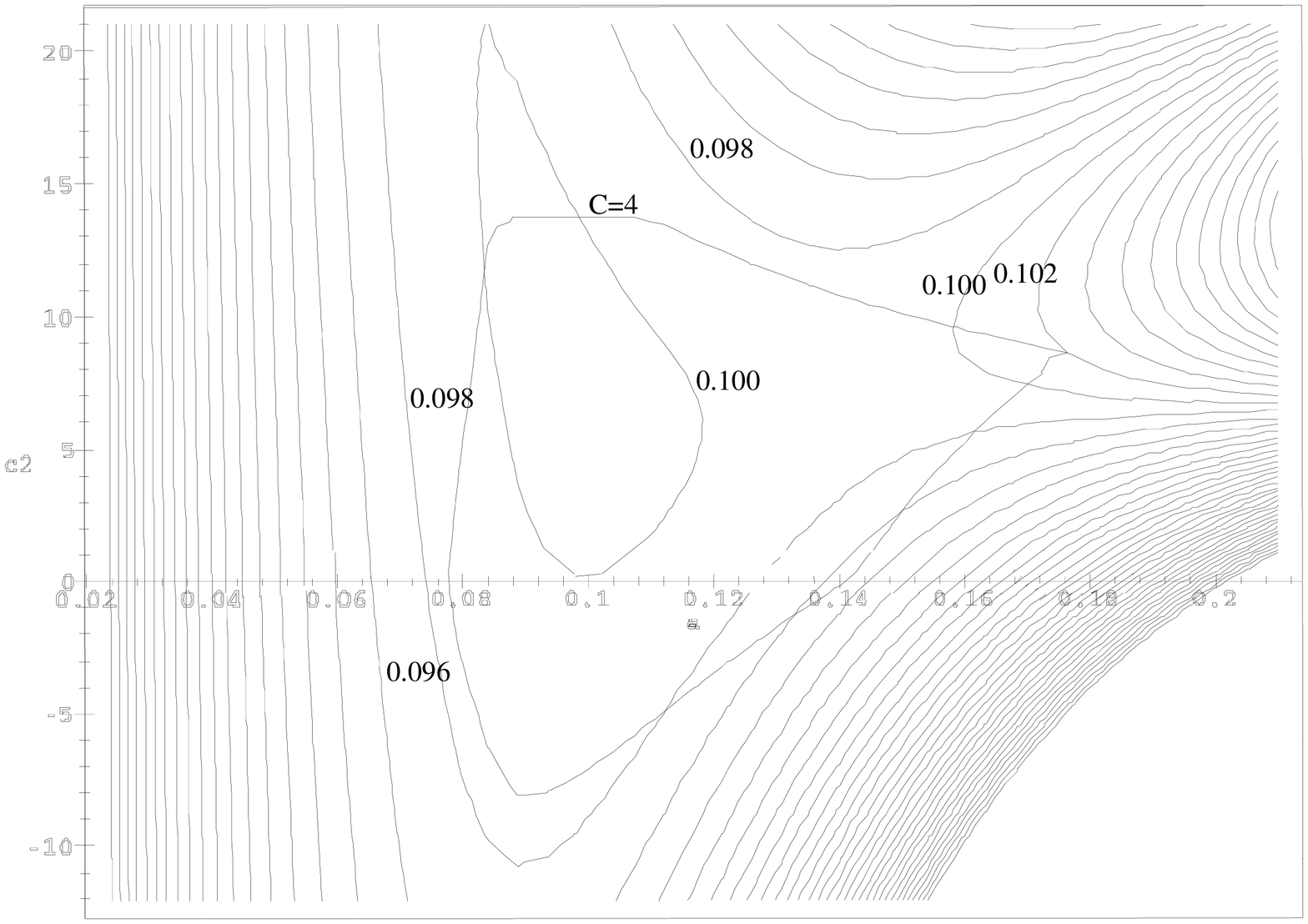,width=17.0truecm,angle=90}}
    \caption{
A contour plot of the RS dependence of the Bjorken Sum Rule
effective charge $a^*_{Bj}$ calculated using a fourth-order partial sum
evaluated with the PMS/ECH fourth-order coefficient. The $a^*_{Bj}$
contours are spaced as in Fig.~1, and we also show the $\CI=4$ contour.}
  \label{PMS improvement}
  \end{center}
\end{figure}

\begin{figure}[htb]
  \begin{center}
\mbox{\kern-0.5cm
\epsfig{file=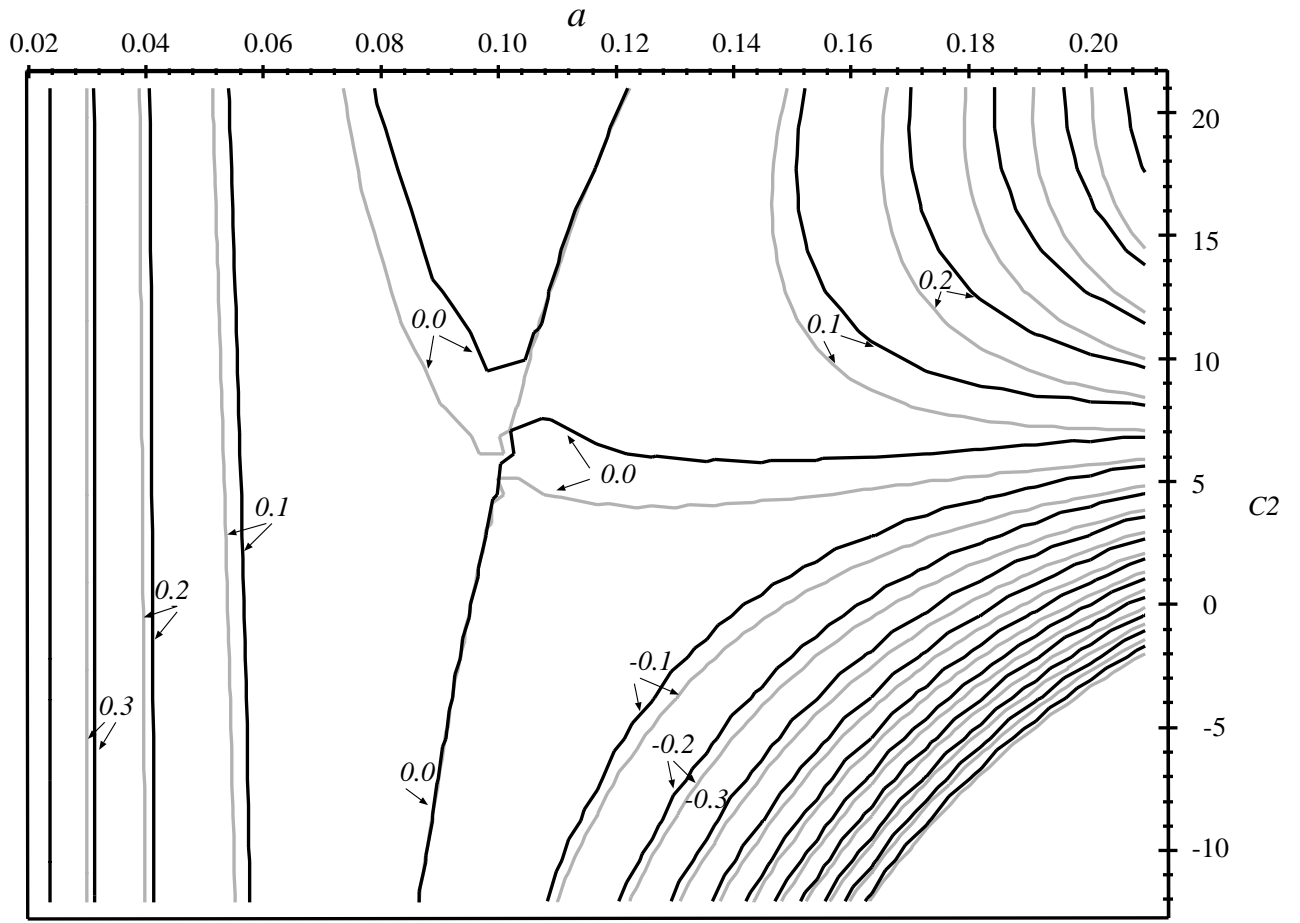,width=17.0truecm,angle=-90}}
    \caption{
The fourth term of the Bjorken Sum Rule series $r_3^{est}a^3$
as predicted by the [0/2] PAP (gray line) and the PMS/ECH (black line).
The separation between each pair of adjacent contours is 0.1.}
\label{PAP_vs_PMS}
  \end{center}
\end{figure}
\vskip 14pt
\end{document}